%% file: main.tex
\documentclass[runningheads]{llncs}
\usepackage{graphicx}
\usepackage{amsfonts}
\usepackage{xcolor}
\newcommand{\printfnsymbol}[1][\value{footnote}]{\footnotemark[#1]}
\usepackage{authblk}
\begin{document}
\title{Explaining COVID-19 and Thoracic Pathology Model Predictions by Identifying Informative Input Features}
\titlerunning{ }
%
\author{Ashkan Khakzar\textsuperscript{1,}\thanks{denotes joint first author, and $^{\ast}$ S. T. Kim and N. Navab shared senior authorship. \newline $^{\dagger}$  denotes corresponding author (st.kim@khu.ac.kr)}
, Yang Zhang\textsuperscript{1,}\printfnsymbol, Wejdene Mansour\textsuperscript{1}, Yuezhi Cai\textsuperscript{1},
Yawei Li\textsuperscript{1}, Yucheng Zhang\textsuperscript{1}, Seong Tae Kim\textsuperscript{2,${\ast}$,${\dagger}$}, Nassir Navab\textsuperscript{1,3,${\ast}$}}
%
%
\authorrunning{ }
%
\institute{\textsuperscript{1}Technical University of Munich\\
\textsuperscript{2}Kyung Hee University\\
\textsuperscript{3}Johns Hopkins University 
}
\maketitle              
\begin{abstract}
\input{Content/abstract}
 
\end{abstract}
\section{Introduction}
\input{Content/introduction}

\section{Methodology}
\input{Content/method}

\section{Experiments and Results}
\input{Content/experiments}

\section{Discussion}
\input{Content/discussion}

\section{Conclusion}
In this work, we build on IBA feature attribution method and come up with different approaches for identifying input regions that have predictive information. Contrary to IBA, our Inverse IBA method identifies \emph{all} regions that can have predictive information. Thus all predictive cues from the pathologies in the X-rays are highlighted. Moreover, we propose Regression IBA for attribution on regression models. In addition, we propose Multi-layer IBA, an approach for obtaining fine-grained attributions which can identify subtle features.

\subsubsection{Acknowledgement}
This work was partially funded by the \textbf{Munich Center for Machine Learning (MCML)} and the \textbf{Bavarian Research Foundatio}n grant AZ-1429-20C. The computational resources for the study are provided by the \textbf{Amazon Web Services Diagnostic Development Initiative}.
S.T. Kim is supported by the \textbf{Korean MSIT}, under the National Program for Excellence in SW (2017-0-00093), supervised by the IITP.

%
%
%
\bibliographystyle{splncs04}
\bibliography{references}

\end{document}

%% file: Content/abstract.tex
Neural networks have demonstrated remarkable performance in classification and regression tasks on chest X-rays.  
In order to establish trust in the clinical routine, the networks' prediction mechanism needs to be interpretable.  
One principal approach to interpretation is feature attribution. Feature attribution methods identify the importance of input features for the output prediction.  
Building on Information Bottleneck Attribution (IBA) method, for each prediction we identify the chest X-ray regions that have high mutual information with the network's output. 
Original IBA identifies input regions that have \emph{sufficient} predictive information.
We propose Inverse IBA to identify \emph{all} informative regions. Thus all predictive cues for pathologies are highlighted on the X-rays, a desirable property for chest X-ray diagnosis.
Moreover, we propose Regression IBA for explaining regression models. Using Regression IBA we observe that a model trained on cumulative severity score labels implicitly learns the severity of different X-ray regions.
Finally, we propose Multi-layer IBA to generate higher resolution and more detailed attribution/saliency maps.  
We evaluate our methods using both human-centric (ground-truth-based) interpretability metrics, and human-agnostic feature importance metrics on NIH Chest X-ray8 and BrixIA datasets. The code\footnote{\url{https://github.com/CAMP-eXplain-AI/CheXplain-IBA}} is publicly available.

%% file: Content/introduction.tex
Deep Neural Network models are the de facto standard in solving classification and regression problems in medical imaging research. Their prominence is specifically more pronounced in chest X-ray diagnosis problems, due to the availability of large public chest X-ray datasets \cite{wang2017chestx,irvin2019chexpert,rajpurkar2017chexnet,johnson2019mimic}. Chest X-ray is an economical, fast, portable, and accessible diagnostic modality. A modality with the aforementioned properties is specifically advantageous in worldwide pandemic situations such as COVID-19 where access to other modalities such as Computed Tomography (CT) is limited \cite{signoroni2020end,oh2020deep,punn2020automated}. Therefore, diagnostic chest X-ray neural network models can be of great value in large-scale screening of patients worldwide.
 
However, the black-box nature of these models is of concern. It is crucial for their adoption to know whether the model is relying on features relevant to the medical condition. In pursuit of interpretability of chest X-ray models, a class of works focuses on instilling interpretability into the models during optimization \cite{taghanaki2019infomask,khakzar2019learning,singh2021covidscreen}, another class pursues optimization semi-supervised with localization \cite{li2018thoracic}, and another class of works provides post-hoc explanations \cite{wang2017chestx,rajpurkar2017chexnet,karim2020deepcovidexplainer}. Post-hoc explanations have the advantage that they can be applied to any model without changing the objective function. 

One principal method for post-hoc explanation is feature attribution (aka saliency methods), i.e. identifying the importance/relevance of input features for the output prediction \cite{simonyan2013deep,sundararajan2017axiomatic,khakzar2021neural,bach2015pixel,selvaraju2017grad,schulz2020restricting}. 
Feature attribution problem remains largely open to this date, however, many branches of solutions are proposed.
The question is which attribution solution to use. Attributions are evaluated from several perspectives, and one crucial and \emph{necessary} aspect is to evaluate whether the attributed features are indeed important for model prediction, which is done by feature importance metrics \cite{samek2016evaluating,ancona2017towards,petsiuk2018rise}. One \emph{desirable} property is human interpretability of the results, i.e. if the attribution is interpretable for the user. For example, Class Activation Maps (CAM, GradCAM) \cite{zhou2016learning,selvaraju2017grad} being a solid method that is adopted by many chest X-ray model interpretation works, satisfies feature importance metrics. However, it generates attributions that are of low resolution, and while accurately highlighting the important features, they do not highlight these regions with \emph{precision}. Such precision is of more importance in chest X-rays where features are subtle. On the other hand, some other methods (e.g. Guided BackPropagation \cite{springenberg2014striving}, $\alpha1\beta0$ \cite{bach2015pixel}, Excitation Backprop \cite{zhang2018top}) have pixel-level resolution and are human-interpretable, but do not satisfy feature important metrics and some do not explain model behavior \cite{adebayo2018sanity,NEURIPS2019_fe4b8556,khakzar2020rethinking,khakzar2019explaining,nie2018theoretical}.
 
Information Bottleneck Attribution (IBA)~\cite{schulz2020restricting} is a recent method proposed in neural networks literature that satisfies feature importance metrics, is more human-interpretable than established methods such as CAMs~\cite{zhou2016learning}, and is of solid theoretical grounding. The method also visualizes the amount of information each image region provides for the output in terms of bits/pixels, thus its attribution maps (saliency maps) of different inputs are comparable in terms of quantity of the information (bits/pixels). Such properties make IBA a promising candidate for chest X-ray model interpretation. 
 
In this work, we build upon IBA and propose extended methodologies that benefit chest X-ray model interpretations.

\subsection{Contribution Statement}
 
\textbf{Inverse IBA:} The original IBA method finds input regions that have sufficient predictive information. The presence of these features is sufficient for the target prediction. However, if sufficient features are removed, some other features can have predictive information. We propose Inverse IBA to find any region that can have predictive information.
\\
\textbf{Regression IBA:} IBA (and many other methods such as CAMs) is only proposed for classification. We propose Regression IBA and by using it we observe that a model trained on cumulative severity score labels implicitly learns the severity of different X-ray regions.
\\
\textbf{Multi-layer IBA:} We investigate approaches to use the information in layers of all resolutions, to generate high-resolution saliency maps that \emph{precisely} highlight informative regions. Using Multi-layer IBA, for instance, we can precisely highlight subtle regions such as Mass, or we observe that the model is using corner regions to classify Cardiomegaly. 
\\
\textbf{Effect of balanced training:} We also observe that considering data imbalance during training results in learned features being aligned with the pathologies.

%% file: Content/method.tex
\subsubsection{Information Bottleneck for Attribution (IBA)~\cite{schulz2020restricting}}
inserts a bottleneck into an existing network to restrict the flow of information during inference given an input. 
The bottleneck is constructed by adding noise into the feature maps (activations) of a layer.
Let $F$ denote the feature maps at layer $l$, the bottleneck is represented by $Z= \lambda F+(1-\lambda)\epsilon$, 
%
where $\epsilon$ is the noise, the mask $\lambda$ has the same dimension as $F$ and controls the amount of noise added to the signal. Each element in the mask $\lambda_i \in [0,1]$. 
Since the goal is post-hoc explanation for an input $X$, the model weights are fixed and \emph{the mask $\lambda$ is optimized} such that mutual information between the noise-injected activation maps $Z$ and the input $X$ is minimized, while the mutual information between $Z$ and the target $Y$ is maximized:
\begin{equation}
\label{eq:bottleneck}
    \max_{\lambda}I(Y,Z)-\beta I(X,Z)
\end{equation}
%
 
The term $I(X,Z)$ is intractable, thus it is (variationally) approximated by 
%
\begin{equation}
\label{eq:approx}
    I(X,Z) 	\approx \mathcal{L}_{I} = E_{F}[D_{KL}(P(Z|F)||Q(Z))] 
\end{equation}
where $Q(Z) \sim \mathcal{N}(\mu_{F},\sigma_{F})$ ($\mu_{F}$ and $\sigma_{F}$ are the estimated mean and variance of hidden feature $F$ from a batch of data samples). In \cite{schulz2020restricting}, the mutual information $I(Y,Z)$ is replaced by cross entropy loss $L_{CE}$. It is proven $-L_{CE}$ is a lower bound for I(Y,Z) \cite{alemi2016deep}. Minimizing $L_{CE}$ corresponds to maximizing $-L_{CE}$ and thus maximizing the lower bound of I(Y,Z). The objective becomes:
\begin{equation}
\label{eq:IBA_loss}
    \mathcal{L} =  \beta \mathcal{L}_{I} + \mathcal{L}_{CE}
\end{equation}

\subsection{Inverse IBA}

In IBA formulation (Eq. \ref{eq:IBA_loss}), the first term tries to remove as many features as possible (by setting $\lambda=0$) and while the second term tries to keep features (by setting $\lambda=1$) such that mutual information with target $Y$ is kept. Therefore, if only a small region can keep the second term ($\mathcal{L}_{CE}$) minimized (keep the mutual information with target $Y$), the rest of the features are removed (their $\lambda=0$). The identified regions ($\lambda=1$) have \emph{sufficient} predictive information, as their existence is sufficient for the prediction. However, there might exist other regions that have predictive information in the absence of these sufficient regions. From another perspective, IBA is playing a preservation game, which results in preserving features that keep the output close to the target.
 
To find all regions that have predictive information we change the formulation of IBA such that the optimization changes to a deletion game. I.e. deleting the smallest fraction of features such that there is no predictive information for the output anymore after deletion.
In order to change IBA optimization to a deletion game we make two changes: 1) for the second term ($\mathcal{L}_{CE}$) in Eq. \ref{eq:IBA_loss} we use an inverse mask: $Z_{inv}= \lambda\epsilon+(1-\lambda)F$, and denote the new term with $\mathcal{L}_{CE}^{inv}$. 2) we maximize the $\mathcal{L}_{CE}^{inv}$ in order for the optimization to remove predictive features. 
Thus, the objective is:

\begin{equation}
\label{eq:IBA_inverse_loss}
    \mathcal{L}_{inv} =  \beta \mathcal{L}_{I} - \mathcal{L}_{CE}^{inv}
\end{equation}

Minimizing $\mathcal{L}_{I}$ corresponds to the feature map becoming noise (similar to IBA) and pushes $\lambda$ to 0. 
Minimizing $\mathcal{L}_{CE}^{inv}$ (maximizing $\mathcal{L}_{CE}^{inv}$) in Eq. \ref{eq:IBA_inverse_loss} corresponds to removing \emph{all} predictive information. In the $\mathcal{L}_{CE}^{inv}$ term, we use $Z_{inv}$, thus removing features corresponds to pushing the $\lambda$ to 1 (if we instead use $Z$ instead of $Z_{inv}$, $\lambda$ moves to 0, and as $\mathcal{L}_{I}$ also pushes $\lambda$ to 0, we get 0 everywhere). Therefore, $\lambda$ is pushed to 1 for \emph{any} predictive feature, and to 0 for the rest. As such, Inverse IBA identifies \emph{any} predictive feature in the image and not just the \emph{sufficiently} predictive features (examples in Fig. \ref{fig:inverseIBA}). 

%

\subsection{Regression IBA} 

Original IBA is proposed for classification setting. In this section, we discuss several variations of IBA for the regression case. We discuss three different regression objectives: 1) MSE Loss defined as $\mathcal{L}_{MSE}=(\mathbf{\Phi}(Z)-\mathbf{y})^{2}$. MSE loss has the property that if the target score is small, it identifies regions with small brixIA score as informative. Because in this case, the objective is trying to find regions that have information for output to be zero. 2) Regression Maximization (RM) Loss is simply defined as $\mathcal{L}_{RM} = \mathbf{\Phi}(Z)^{2}$. This loss has the property that it favors regions with high scores as informative. 3) Deviation loss defined as $\mathcal{L}_{DV}=(\mathbf{\Phi}(Z)-X)^{2}$. We subtract the score of the noisy feature map from the score of the original image. Similar to IBA for classification, this formulation identifies regions with sufficient information for the prediction. We also apply Inverse IBA to regression (see Fig. \ref{fig:inverseIBA}) to identify all preditive features.
 
\subsection{Multi-layer IBA}

For original IBA, the bottleneck is inserted in one of the later convolutional layers. As we move towards earlier layers, the variational approximation becomes less accurate. Thus the optimization in Eq. \ref{eq:IBA_loss} highlights extra regions that do not have predictive information in addition to highlighting the sufficient regions. However, as the resolution of feature maps in earlier layers are higher, the highlighted regions are crisper and more interpretable. In order to derive regions that are crips and have high predictive information we compute IBA for several layers and combine their results, thus introducing Multi-layer IBA:
\begin{equation}
    \mathcal{T}(IBA_{L_{1}}) \cap  \mathcal{T}(IBA_{L_{2}}) ... \cap \mathcal{T}(IBA_{L_{L}})
\end{equation}
\noindent where $\mathcal{T}$ denotes a thresholding operation to binarize the IBA maps.

\subsection{Chest X-ray Models} 

\textbf{Classification model: } We denote a neural network function by $\mathbf{\Phi}_{\Theta}(\mathbf{x}):\mathbb{R}^{H\times W} \rightarrow \mathbb{R}^{C}$ where $C$ is the number of output classes. For a dataset $\mathbf{X} =\{\mathbf{x}^{(1)}, ..., \mathbf{x}^{(N)}\}$, and their labels $\mathbf{Y}=\{\mathbf{y}^{(1)}, ..., \mathbf{y}^{(N)}\}$, where $\mathbf{y} = [\mathbf{y}_{j}]^C$, and $\mathbf{y_{j}}\in\{0,1\}$. Chest X-rays can have multiple pathologies. We use Binary Cross Entropy (BCE) loss on each output for multilabel prediction.
\begin{equation}
\mathcal{L}_{BCE} = (\hat{\mathbf{y}}, \mathbf{y}) = - \sum_{j} \beta \mathbf{y_{j}}\, \log (\hat{\mathbf{y_{j}}}) + (1-\mathbf{y_{j}})\, \log (1-\hat{\mathbf{y_{j}}}) 
\label{eq:loss_bce}
\end{equation}
where $\beta$ is a weighting factor to balance the positive labels. 
\\ 
\\
\textbf{Regression model:} Consider a neural network $\mathbf{f}_{\Theta}(\mathbf{x}):\mathbb{R}^{H\times W} \rightarrow \mathbb{R}$ and a dataset $\mathbf{X} =\{\mathbf{x}^{(1)}, ..., \mathbf{x}^{(N)}\}$ of $N$ X-ray images, and their corresponding labels $\mathbf{Y}=\{\mathbf{y}^{(1)}, ..., \mathbf{y}^{(N)}\}$, where $\mathbf{y}_{j} \in {0,...,18}$ is the cumulative severity score on each image. We model the regression problem with a MSE loss:
\begin{equation}
    \mathcal{L}_{MSE} = \frac{1}{N} \sum (\mathbf{\mathbf{\Phi}_{\Theta}(\mathbf{x})^{(n)}}-\mathbf{y^{(n)}})^{2}
\end{equation}

%% file: Content/experiments.tex
 
\subsubsection{Implementation Details}
We use three models: 1) NIH ChestX-ray8 classification: Network with 8 outputs for the 8 pathologies. 2) BrixIA regression: Network with one output and predicts the total severity score (sum of severity scores of 6 regions) 3) BrixIA classifier: 3 outputs detecting whether a severity score of 3, 2, and 0/1 exists in the X-rays.
We use Densenet 121, and insert the IBA bottleneck on the output of DenseBlock 3. For Multi-layer IBA we insert it on the outputs of DenseBlock 1,2 and 3.
%
%
\begin{figure}[t]
\includegraphics[width=\textwidth]{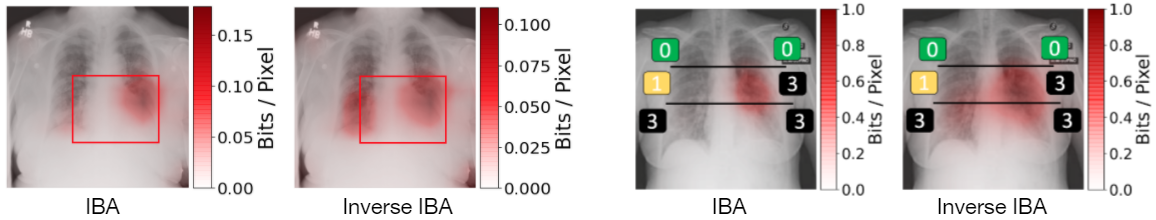}
\caption{\textbf{Inverse IBA:} Inverse IBA compared with IBA on a sample from the NIH Chest X-ray8 (left) and a sample from BrixIA (right). \textbf{NIH Chest X-ray8 (left)}: Inverse IBA is identifying both sides of Cardiomegaly as informative. The bounding box denotes the expert's annotation. \textbf{BrixIA (Right)}: IBA is identifying two regions with a severity score of 3 as sufficient for predicting the score of 3, however, Inverse IBA is identifying all regions with a severity score of 3, and the region with a score of 1. The horizontal lines denote the 6 regions within the lungs, and the numbers represent the sevirity score of each region.}
\label{fig:inverseIBA}
\end{figure}
 
\subsection{Feature Importance (Human-Agnostic) Evaluations}
\begin{figure}[t]
\includegraphics[width=\textwidth]{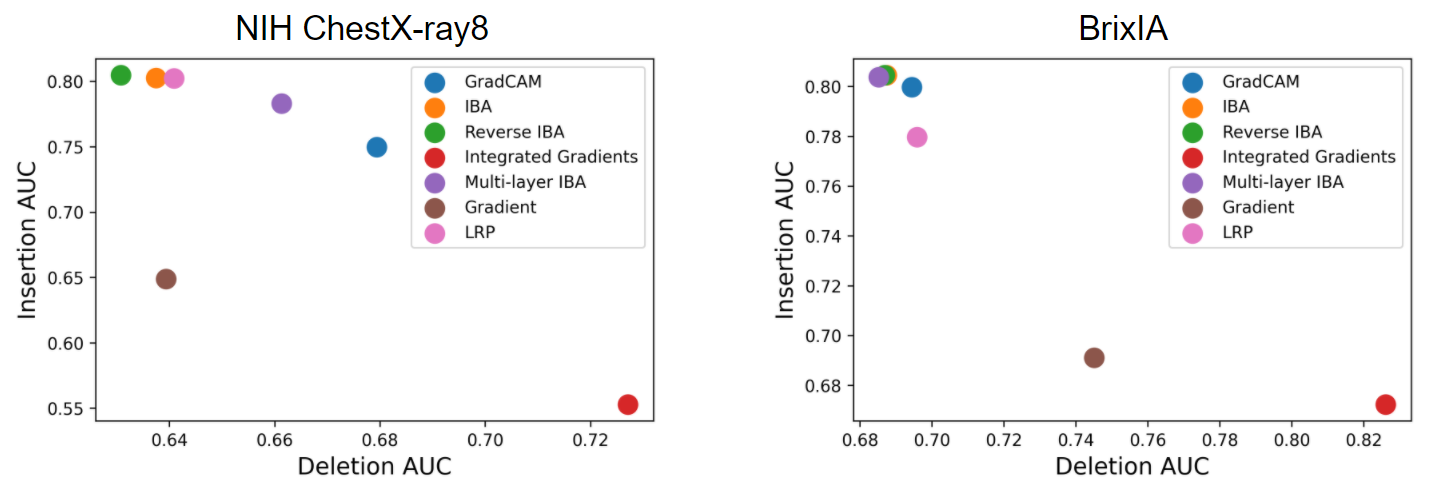}
\caption{\textbf{Insertion/Deletion metric:} Comparison of different attribution methods in terms of feature importance. Method with high Insertion AUC and low Deletion AUC is the best (top left corner is the best).} \label{fig:insertion}
\end{figure}

Experiments in this section evaluate whether an attribution method is identifying important features for the model prediction.
\\
\textbf{Insertion/Deletion~\cite{samek2016evaluating,petsiuk2018rise}}
Insertion: given a baseline image (we use the blurred image) features of the original image are added to the baseline image starting from the most important feature and the output is observed. If the attribution method is correct, after inserting a few features the prediction changes significantly, thus the AUC of output is high. Deletion: deleting important features first. The lower the AUC the better. Results are presented in Fig. \ref{fig:insertion}.
%
\begin{figure}[t]
\includegraphics[width=\textwidth]{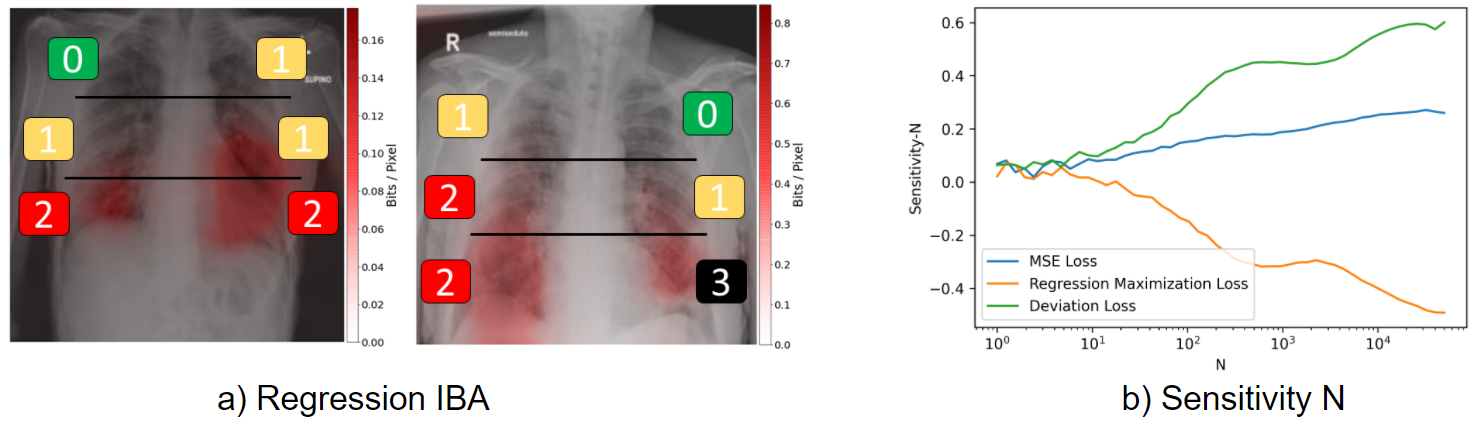}
\caption{\textbf{Regression IBA:} a) Regression IBA (with $\mathcal{L}_{DV}$) applied on a regression model that predicts the total severity score. Using Regression IBA we see that the model has identified the severity scores of different regions implicitly. b) Sensitivity N metric for evaluating feature importance of different Regression IBA losses} \label{fig:regression}
\end{figure}
\\
\textbf{Sensitivity-N~\cite{ancona2017towards}}
We adapt this metric to regression case (not trivial with Insertion/Deletion) and use it to evaluate Regression-IBA. 
In Sensitivity-n we mask the input randomly and observe the correlation between the output change and the values of attribution map overlapping with the mask. The higher the correlation the more accurate the attribution map. Results in Fig. \ref{fig:regression}b.
%
 
\subsection{Ground Truth based (Human-centric) Evaluations}
Experiments in this section evaluate the attribution maps in terms of the human notion of interpretability, i.e. the alignment between what we understand and the map. Moreover, they measure how fine-grained the attribution maps are.
\\
\\
\textbf{Localization} For NIH ChestX-ray8 dataset the bounding boxes are available. To generate boxes for BrixIA score regions, similar to \cite{signoroni2021bs} we use a lung segmentation network with the same architecture in \cite{signoroni2021bs}. We divide each lung into 3 regions. We threshold the attribution maps and compute their IoU with these divided regions (over dataset).
%
\begin{table}[t]
\caption{Mean IOU on NIH ChestX-ray8 (all pathologies) for various methods}
\label{tab:NIH_attribution_methods}
\centering
\begin{tabular}{|l|l|l|l|l||l||l|}
\hline
 GradCAM\cite{selvaraju2017grad} &  InteGrad\cite{sundararajan2017axiomatic} & LRP\cite{montavon2017explaining} & Gradients\cite{simonyan2013deep} & IBA & Inverse IBA & Multi-layer IBA\\
\hline
0.077 & 0.076 & 0.025 & 0.114 & 0.114 & 0.088 & \textbf{0.189} \\
\hline
\end{tabular}
\end{table}
\begin{table}[t]
\caption{Mean IOU on BrixIA for each detector and for various attribution methods}
\label{tab:BrixIA_attribution_methods}
\centering
\begin{tabular}{|l|l|l|l|l|l||l||l|}
\hline
 & GradCAM &  InteGrad& LRP & Gradients& IBA & Inverse IBA & Multi-layer IBA\\
\hline
 Detector 0/1& 0.11 & 0.176 & 0.0 & 0.04 & 0.145 & \textbf{0.194} & 0.171  \\
\hline
Detector 2& 0.0 & 0.13 & 0.0 & 0.019 & 0.13 & 0.245 & \textbf{0.257}  \\
\hline
Detector 3& 0.011 & 0.14 & 0.0 & 0.052 & 0.222 & 0.243 & \textbf{0.257}  \\
\hline
\end{tabular}
\end{table}
\\
\noindent\textbf{Correlation Analysis (Regression Models)}
For the BrixIA dataset, we evaluate the performance of regression models by measuring the correlation between the attribution scores and the severity scores. For each image, we first assign each pixel with its severity score, obtaining a severity score map. We then flatten both the attribution and severity score maps and compute their Pearson correlation coefficient (PCC). The PCC results are as follows:  for $\mathcal{L}_{MSE}$, 0.4766, for $\mathcal{L}_{RM}$, 0.4766, for $\mathcal{L}_{MSE}$, 0.4404, and for random heatmaps, 0.0004.
\begin{table}[h]
\caption{Mean IOU on NIH ChestX-ray8 dataset for BCE and Weighted BCE models, reported for all the pathologies in NIH Chest X-ray8 using Inverse IBA}
\label{tab:IOU_wBCE}
\centering
\begin{tabular}{|l|l|l|l|l|l|l|l|l||l|}
\hline
 &  Atelec. & Cardio. & Effusion & Infiltrate. & Mass & Nodule & Pneumo. & Pn. thorax & Mean\\
\hline
BCE & 0.016 & 0.071 & 0.004 & 0.001 & 0.102 & 0.011 & 0.0 & 0.003 & 0.024\\
W. BCE & 0.073 & 0.131 & 0.032 & 0.058 & 0.097 & 0.02 & 0.066 & 0.016 & \textbf{0.065}\\
\hline
\end{tabular}
\end{table}

\begin{figure}[!t]
\includegraphics[width=\textwidth]{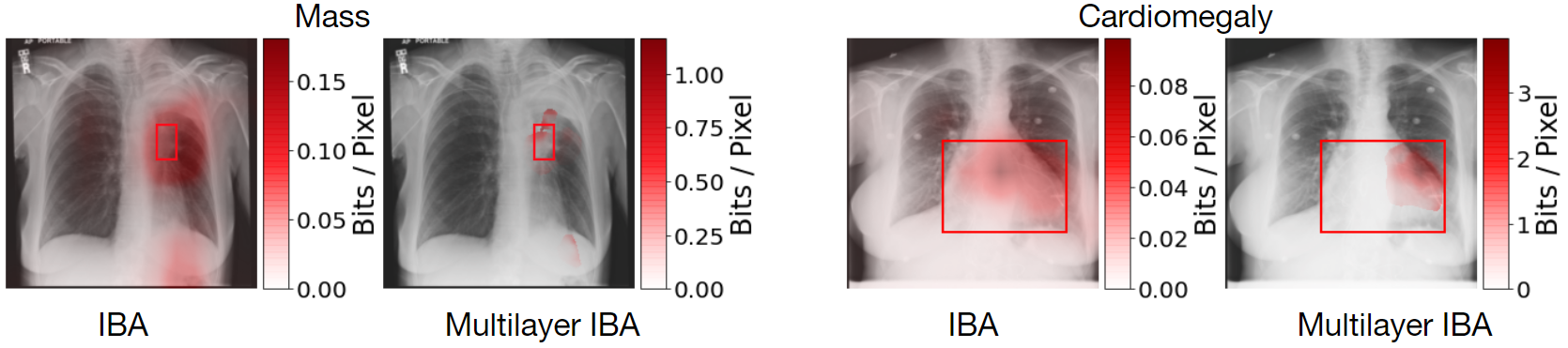}
\caption{\textbf{Multi-layer IBA:} Multi-layer IBA generates more fine-grained maps compared to IBA. (Left): Multi-layer IBA precisely highlights subtle features such as Mass (Right) Using Multi-layer we observe that for Cardiomegaly in this X-ray the corner regions of Cardiomegaly are used. IBA highlights the entire region.}
\label{fig:multilayer}
\end{figure}

%% file: Content/discussion.tex
\subsubsection{Inverse IBA:}
We observe that (Fig. \ref{fig:inverseIBA}) Inverse IBA highlights all regions with predictive information. On BrixIA sample, IBA only identifies two regions with a score of 3 as being predictive, while Inverse IBA identifies all regions with a score of 3. On NIH sample, if we remove the highlighted areas of both methods (equally remove) from the image, the output change caused by the removal of Inverse IBA regions is higher. This is also quantitatively validated across dataset in the Deletion experiment (Fig. \ref{fig:insertion}).
\\
\textbf{Regression IBA:}
Using Regression IBA we observe that (Fig. \ref{fig:regression}) a regression model which only predicts one cumulative severity score (0-18) for each X-ray implicitly identifies the severity scores of different regions. 
\\
\textbf{Multi-layer IBA:}
We use Multi-layer IBA for obtaining fine-grained attributions. In Fig. \ref{fig:multilayer} we see that such fine-grained attributions allow for identifying subtle features such as Mass. Moreover, Multi-layer IBA also uncovers some hidden insights regarding what features the model is using for the Cardiomegaly example. While IBA highlights the entire region, Multi-layer IBA shows precisely the regions to which IBA is pointing.
\\
\textbf{Imbalanced loss:}
We observe in Tab. \ref{tab:IOU_wBCE} that using weighted BCE results in an increased IoU with the pathologies. This signifies that the contributing features of the weighted BCE model are more aligned with the pathology annotations. The observation is more significant when we consider the AUC of ROC and the Average Precision (AP) of these models. The AUCs (BCE=0.790, Weighted BCE=0.788) and APs (BCE=0.243, Weighted BCE=0.236) are approximately equivalent. The BCE even archives marginally higher scores in terms of AUC of ROC and AP but its learned features are less relevant to the pathologies.